\documentclass[aip,reprint]{revtex4-1}

\draft % marks overfull lines with a black rule on the right

\usepackage{graphicx}% Include figure files
\usepackage{hyperref}
\usepackage{dcolumn}% Align table columns on decimal point
\usepackage{bm}% bold math

\usepackage[utf8]{inputenc}
\usepackage[T1]{fontenc}
\usepackage{mathptmx}
\usepackage{etoolbox}

\usepackage{physics}
\usepackage{amsmath}

\begin{document}

%\preprint{}

\title{A cold-atom Ramsey clock with a low volume physics package} %Title of paper



\author{A. Bregazzi}
 \affiliation{SUPA and Department of Physics, University of Strathclyde, G4 0NG, Glasgow, United Kingdom}

\author{E. Batori}
\affiliation{University of Neuch\^atel, Institute of Physics, Laboratoire Temps-Fr\'equence, Avenue de Bellevaux 51, 2000
Neuch\^atel, Switzerland}

\author{B. Lewis}
 \affiliation{SUPA and Department of Physics, University of Strathclyde, G4 0NG, Glasgow, United Kingdom}

\author{C. Affolderbach}
\affiliation{University of Neuch\^atel, Institute of Physics, Laboratoire Temps-Fr\'equence, Avenue de Bellevaux 51, 2000
Neuch\^atel, Switzerland}

\author{G. Mileti}
\affiliation{University of Neuch\^atel, Institute of Physics, Laboratoire Temps-Fr\'equence, Avenue de Bellevaux 51, 2000
Neuch\^atel, Switzerland}

\author{E. Riis}
 \affiliation{SUPA and Department of Physics, University of Strathclyde, G4 0NG, Glasgow, United Kingdom}

\author{P. F. Griffin}
 \email{paul.griffin@strath.ac.uk}
 \affiliation{SUPA and Department of Physics, University of Strathclyde, G4 0NG, Glasgow, United Kingdom}

%Not sure on authorship order so most are left out for now


\date{\today}

\begin{abstract}

We demonstrate a Ramsey-type microwave clock interrogating the 6.835~GHz ground-state transition in cold \textsuperscript{87}Rb atoms loaded from a grating magneto-optical trap (GMOT) enclosed in an additively manufactured loop-gap resonator microwave cavity. A short-term stability of $1.5 \times10^{-11} $~$\tau^{-1/2}$ is demonstrated, in reasonable agreement with predictions from the signal-to-noise ratio of the measured Ramsey fringes. The cavity-grating package has a volume of $\approx$67~cm\textsuperscript{3}, ensuring an inherently compact system while the use of a GMOT drastically simplifies the optical requirements for laser cooled atoms. This work is another step towards the realisation of highly compact portable cold-atom frequency standards.

\end{abstract}

\pacs{}% insert suggested PACS numbers in braces on next line

\maketitle %\maketitle must follow title, authors, abstract and \pacs


%\section{Contents}

%Figures/contents:
%\begin{enumerate}
%    \item Experimental set-up/ sequence
%    \item State prep(?) may be nice to show Zeeman spectroscopy for good state prep and field orientation factor
%    \item Ramsey fringes
%    \item SNR vs Ramsey time study. Would lead into ADEV and show that the cavity launch is a good route to getting better stabilities. Maybe Zeeman scan stuff would% have to drop out to make room??
%    \item ADEV
%\end{enumerate}

%\section{Introduction}

%Intro material, why compact clocks are great, why GMOTS and LGR cavities are also great....

Compact frequency standards based on the interrogation of both ions and neutral atoms continue to receive much interest, with many different schemes now reported in the literature \cite{CSAC_Knappe2004,POP_Micalizio2012,NIST_cold_CPT_Esnault2013,high_performance_POP_Kang2015,Hg_ion_Tjoelker2016,Ion_clock_Cao2017,cold_atom_CPT_Elvin2019,NIST_2photon_Maurice2020,CPT_Ramsey_Carle2021}. While compact clocks based on thermal atomic vapours and coherent population trapping (CPT) remain unrivalled in terms of size, weight and power (SWaP) they typically exhibit stabilities that are limited in the medium to long-term due to light-shift effects \cite{CPT_light_shift_AbdelHafiz2018} and the buffer gasses required to reduce atomic collisions \cite{CSAC_Knappe2004}. In addition to this, CPT clocks struggle to reach atomic shot noise due to the limited signal-to-noise ratio (SNR), a consequence of the low detected signal photon count per atom. This limitation often requires complex interrogation schemes to optimise the clock stability \cite{push_pull_OP_Liu2013,high_contrast_CPT_Yun2016,auto_balanced_Ramsey_CPT_Yudin2018}.

The development in the last decade of pulsed optically pumped (POP) clocks in thermal vapours has driven new research, with state-of-the-art stabilities \cite{POP_Micalizio2012,LGR_Stefanucci2012,LGR_Hao2016}. However, these clocks still suffer from buffer gas shifts, providing the ultimate limitation to their long-term stabilities. Achievable Ramsey times within these systems are also limited to a few ms by spin relaxation of the thermal atoms \cite{POP_Micalizio2012,relaxation_vapour_cell_Batori2022}, restricting the short-term stability performance.

In an effort to combat these limitations several groups have now developed compact cold atom microwave clocks based on spherical optical-integrating sphere cavities \cite{HORACE_Esnault2010,HORACE2_Esnault2011}, cylindrical cavities \cite{cylindrical_cold_atom_clock_Mller2011,cylindrical_cold_atom_clock_Liu2015} and more recently loop-gap-resonator cavities \cite{cold_Atom_cavity_Lee2021}. Three examples of cold atom microwave clocks are now even commercially available \cite{MuClock,Spectra_dynamics,AO_sense}. Different laser cooling schemes with varying optical geometries have been used within these systems, with isotropic cooling of an optical molasses \cite{HORACE_Esnault2010,HORACE2_Esnault2011,cylindrical_cold_atom_clock_Liu2015,MuClock}, pyramid MOTs \cite{israeli_Cavity_clock_Givon2022} and larger traditional 6-beam MOTs \cite{cylindrical_cold_atom_clock_Mller2011,cold_Atom_cavity_Lee2021} all being utilised.

 \begin{figure}[t]
 \includegraphics[width=0.49 \textwidth]{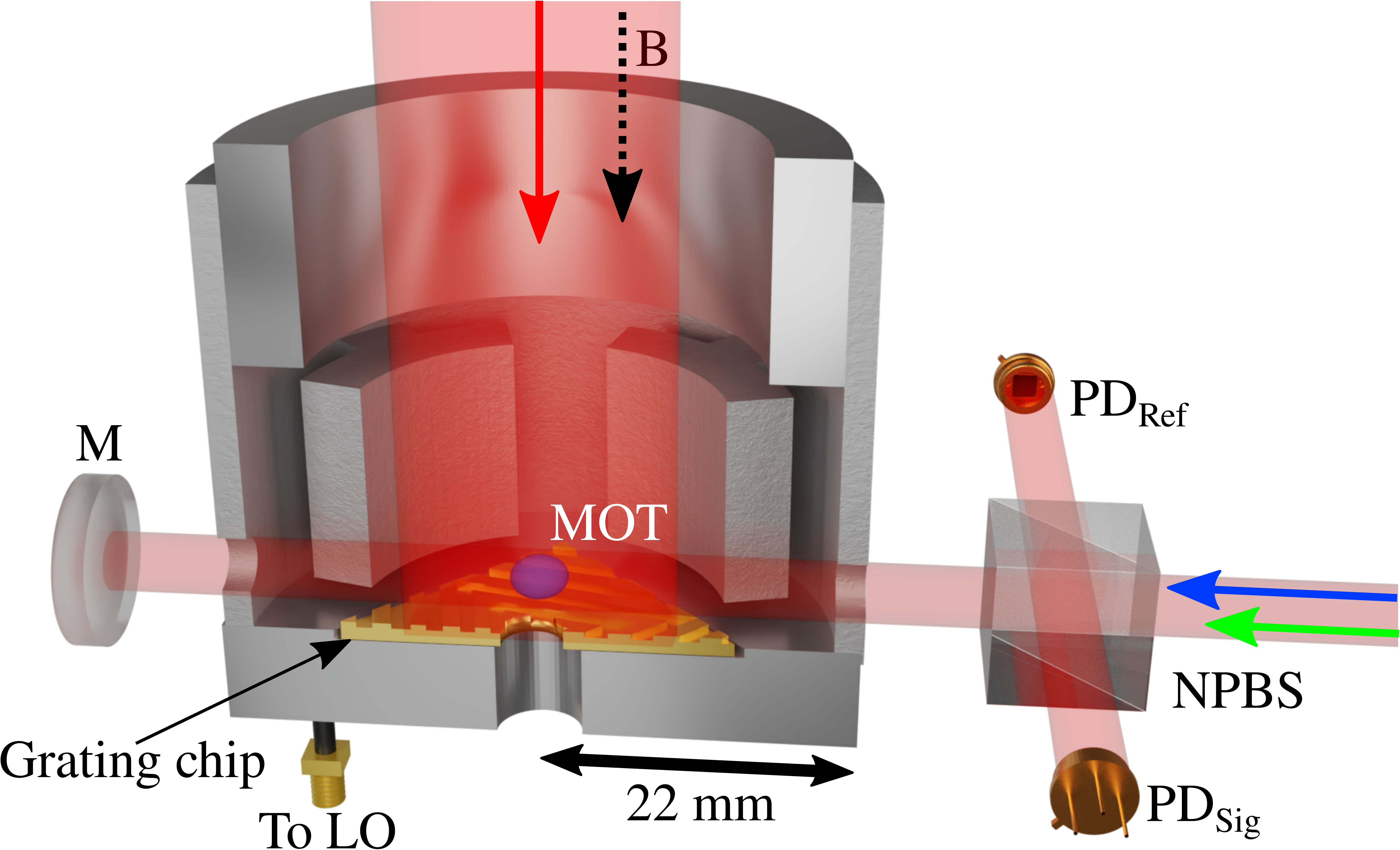}
 \caption{Simplified schematic of the physics package. Trap light (red arrow) propagates parallel to the magnetic bias field (black dashed arrow), 90:10 non-polarising beam splitter (NPBS) splits the optical pumping (blue arrow) and readout (green arrow) light onto a reference photodiode (PD\textsubscript{Ref}) and signal photodiode (PD\textsubscript{Sig}) after being retro-reflected by a mirror (M). Local oscillator (LO) is connected via a SMA vacuum feedthrough.}
 \label{fig:exp_setup}
 \end{figure}

In this letter we demonstrate operation and first stability measurement of a cold-atom atomic clock based on an additively manufactured loop-gap resonator (LGR) cavity with a volume of $\approx67$~cm\textsuperscript{3} incorporating an integrated GMOT \cite{Nshii2013}. Although vapour-cell atomic clocks based on LGR structures with external diameters of order 1~cm have been demonstrated \cite{relaxation_vapour_cell_Batori2022}, such small devices will have limited benefit when using cold atoms. Previous investigations with both GMOTs\cite{Simple_imaging_Bregazzi2021} and 6-beam MOTs\cite{MOT_beam_diameter_Hoth2013} have shown that by constraining the trap beam diameter the maximum trapped atom number is limited, therefore constricting the potential stability of the final clock.

In the present study a sample of cold atoms is optically pumped into the clock state, probed in a double resonance Ramsey-type scheme and the resulting state populations read out through an absorptive method. The use of additive manufacturing allows for complex electrode geometries, while maintaining a highly scalable manufacturing process due to the lack of precision machining and assembly required \cite{LGR_3Dprint_Affolderbach2018,Cavity_paper_Batori2023}. While the main aim of this work has been to demonstrate the feasibility of integrating this cavity design and its fabrication technique with the GMOT architecture, we envisage that further substantial reduction in size is possible by adapting the cavity to form the bulk of the vacuum system. The minimal optical access required of this scheme also allows for a reduction in the number of apertures present in the cavity body which must be carefully considered during the design process so as not to degrade the cavity mode and field homogeneity \cite{cold_Atom_cavity_Lee2021,Cavity_paper_Batori2023}.

%\section{Experiment}

A commercial laser system (Muquans ILS) at 780~nm with an integrated electro-optic modulator (EOM) for the creation of optical sidebands on the carrier frequency is used throughout. The laser light is split into three distinct optical paths, each with a double passed acousto-optic modulator (AOM) for power and frequency control to enable trapping, optical pumping and state readout. The trapping light is coupled into a fibre to be passed to the physics package. The optical pumping and state readout beams are coupled into a single additional fibre and likewise passed to the physics package.

The microwave cavity itself, described in detail in Ref\cite{Cavity_paper_Batori2023} consists of a loop-gap structure with a four electrode geometry and has an outer radius of 44~mm and a height of roughly 44~mm. The cavity operates in a TE\textsubscript{011}-like mode, tuned to the ground-state-splitting of \textsuperscript{87}Rb with a quality factor of Q$\approx$360 and is mounted within a stainless steel vacuum chamber, maintained at ultra-high-vacuum (UHV) by an ion pump. Optical access is enabled via viewports in the vacuum chamber along two orthogonal axes. The first axis is parallel to the cylindrical symmetry axis of the cavity (in the following this is referred to as the cavity axis) and allows the trap light to be directed onto the grating chip after expansion and collimation from the fibre. Optical pumping and state readout light, expanded from the fibre to a 1/e\textsuperscript{2} diameter of 7~mm, is directed onto the atoms through two 4~mm holes drilled in the side of the cavity body. A retro-reflecting mirror for this light is placed outside the vacuum chamber to decrease the acceleration experienced by the atoms when interacting with the optical pumping and probe beams and increase signal amplitudes. A simplified schematic of this is shown in Fig.\ref{fig:exp_setup}. No significant degradation of the cavity field is observed by the introduction of the holes in the cavity body\cite{Cavity_paper_Batori2023}.

A pair of anti-Helmholtz coils are mounted within the vacuum chamber, along the cavity axis in order to create the quadrupole magnetic field required for the trapping process. Three orthogonal pairs of Helmholtz shim coils, mounted externally to the vacuum chamber are used for the cancellation of external stray DC magnetic fields and to apply a magnetic bias along the cavity axis of $\approx100$~mG in order to lift the atomic degeneracy during optical molasses and clock interrogation. We note that the current demonstration is a proof of concept with no magnetic shielding of the experiment present, limiting the potential stability of the system.

The experimental cycle is initiated by turning the trapping coils on with the trap light tuned to be approximately $\Delta$=-2$\Gamma$ red detuned ($\Gamma/2\pi$=6.07~MHz) from the \textsuperscript{87}Rb D\textsubscript{2} \textit{F}=2$\rightarrow$\textit{F'}=3 cycling transition, with re-pump light generated by the EOM operating at 6.57~GHz to produce 5\% optical sidebands. A Rb vapour, maintained at the $1\times10^{-9}$~Torr level, is produced by resistively heating a pair of alkali metal dispensers. We then perform a 6~ms optical molasses by turning the trap coils off and linearly ramping the light detuning to $\Delta$=-5$\Gamma$ while simultaneously decreasing the trap light intensity. After molasses we measure atomic temperatures of $\approx10~\mu K$. In order to decrease the clock cycle time and mitigate experimental dead time, we employ atom recapture between experimental cycles \cite{NIST_cold_CPT_Esnault2013,recapture_Rakholia2014}. In steady state this allows the trapping of $>3\times10^6$ atoms with a load time of 100~ms for a clock cycle operating at $\approx$7~Hz. Once the atoms have been trapped and cooled, the trap light is extinguished by an AOM and blocked by a mechanical shutter \cite{shutters_Zhang2015} to ensure no light leakage during microwave interrogation. After molasses, the atoms are assumed to be roughly evenly distributed between the five $m_F$ levels of the $F=2$ hyperfine ground-state manifold. We therefore employ a 1~ms optical pumping stage with 10~$\mu$W total power of linearly polarised light, polarisation axis parallel to the quantization axis, tuned to the \textit{F}=2$\rightarrow$\textit{F'}=2 and re-pump light tuned to the \textit{F}=1$\rightarrow$\textit{F'}=2 transition. Due to selection rules this pumps $>95\%$ of the atoms into the 5\textsuperscript{2}S\textsubscript{1/2}, 
$\ket{{F=2,m_F=0}}$ state \cite{Optical_pumping_Duan2017} as seen by the almost complete elimination of the $m_F=1\rightarrow m_F'=1$ transitions, when scanning the microwave detuning, increasing the contrast of the resulting clock signal.

 \begin{figure}
 \includegraphics[width=0.49 \textwidth]{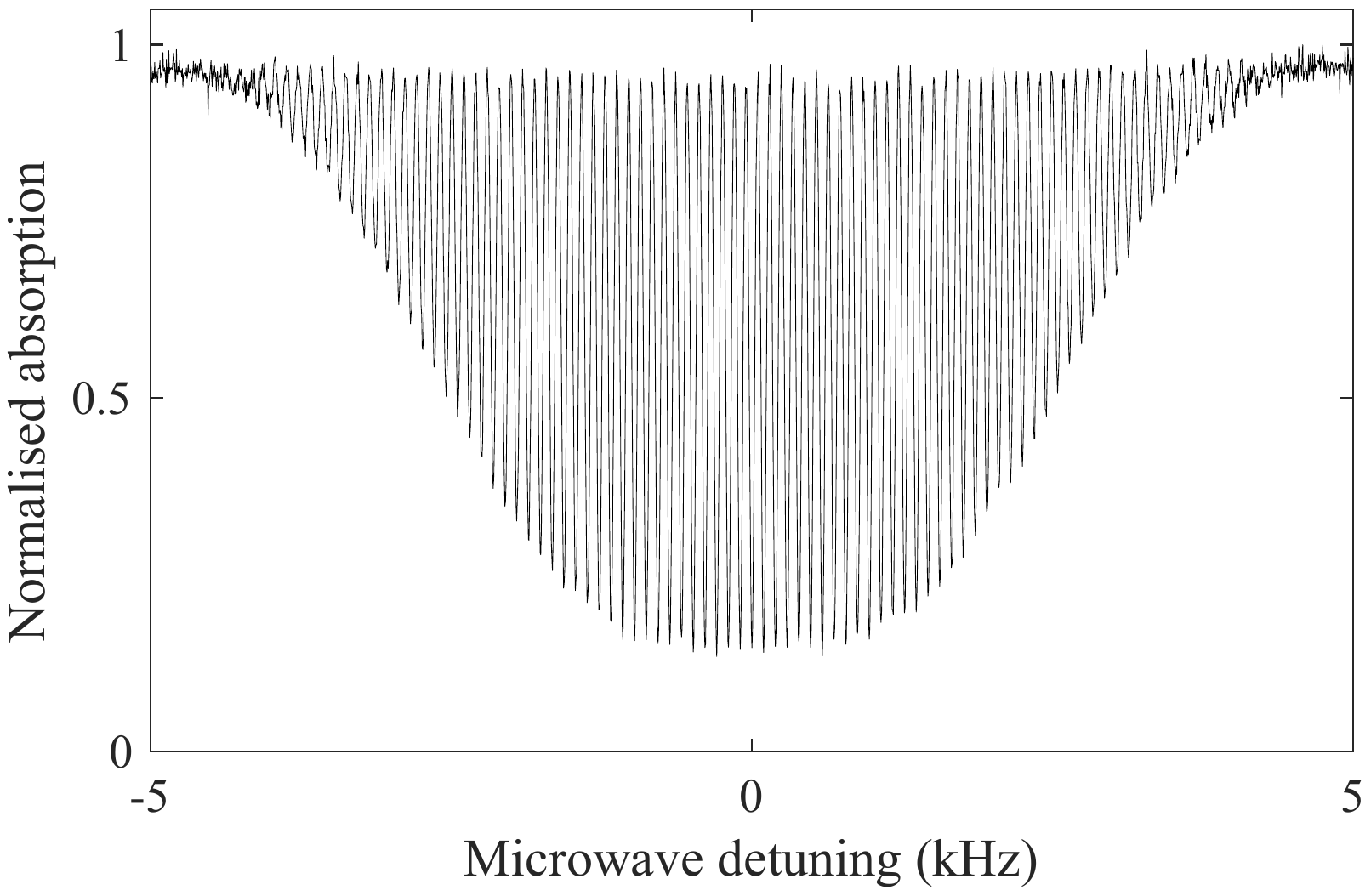}%
 \caption{Typical Ramsey fringe taken with a 10~ms free evolution time.}%
 \label{fig:Ramsey_fringe}
 \end{figure}

A Keysight E8257D microwave synthesizer is used as a local oscillator and microwave source throughout. Square microwave $\pi/2$ pulses with a duration of 200~$\mu$s and typical power levels of 0.04~mW are applied to the atoms. The specified phase noise performance of the local oscillator allows the Dick effect \cite{Dick_effect_Santarelli1998} to be reduced to the $6.12\times10^{-13}$ level at 1~s for the experimental cycle time. After the desired microwave pulses have been applied, the atomic states are read out by an absorption method using 30~$\mu$W of optical power, measured by recording the transmission on a photodiode of two subsequent probe pulses. First a short pulse of readout light tuned to the \textit{F}=2$\rightarrow$\textit{F'}=3 transition is applied, giving an measure of the number of atoms in the $F=2$ ground-state. We then apply a second readout pulse with re-pump light present. This measures the total number of atoms, providing a normalisation of the signal and reduced sensitivity to atom number fluctuations. Intensity noise in the readout pulses is reduced by the use of a reference photodiode.

%To demonstrate efficient optical pumping and field orientation of the microwave field with respect to the magnetic bias field we first preform Zeeman spectroscopy on the hyperfine ground-state of the atoms where a $\pi$-pulse is applied to the atoms and the microwave detuning is scanned. An example scan is shown in Fig.~\ref{fig:Zeeman_spec}. From Fig.~\ref{fig:Zeeman_spec} we see one large central peak (corresponding to the [$F=2, m_F=0] \rightarrow [F=1,m_F=0$] transition) near zero detuning with a characteristic $sinc^2$ distribution due to the square microwave pulse applied and no other peaks outwith the baseline noise on the data. This indicates excellent state preparation of almost all the atoms. We can also infer from these data a good microwave field orientation factor as no $\Delta m_F\neq0$ transitions are visible.

\begin{figure}
 \includegraphics[width=0.49 \textwidth]{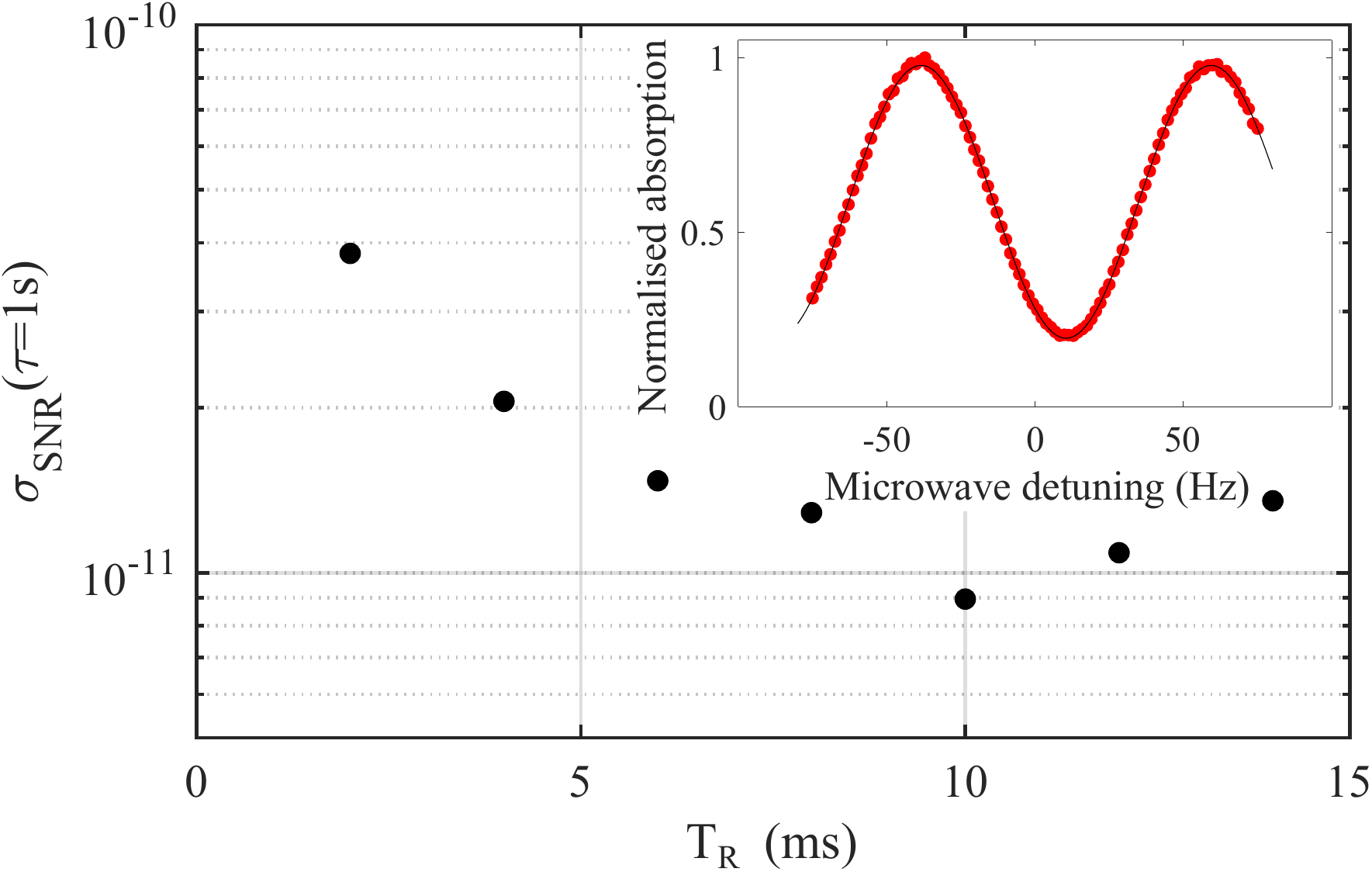}%
 \caption{Predicted short-term stability as a function of Ramsey time. Inset: Typical central Ramsey fringe obtained at 10~ms free evolution time. Red points indicate measured normalised probe absorption while the black line shows a sinusoidal fit to the data. The fringe offset from 0~Hz detuning corresponds to the frequency shift expected from the second-order Zeeman shift at a 100~mG bias field.}
 \label{fig:SNR_optimisation}
 \end{figure}

An example Ramsey fringe taken with a Ramsey time, $T_R$=10~ms, corresponding to a fringe linewidth of 50~Hz is shown in Fig.\ref{fig:Ramsey_fringe}. This fringe exhibits an SNR of around 110, measured at the half-width points of the central fringe, where SNR is defined as the ratio of the peak amplitude divided by the amplitude noise. The predicted SNR limited short-term relative frequency instability of a local oscillator stabilised to an atomic transition in terms of Allan deviation is given by \cite{Frequency_standards_riehle2006}: 

\begin{equation}
    \sigma_{SNR}(\tau)=\frac{1}{\pi C}\frac{\Delta f}{f_0}\frac{1}{SNR}\sqrt{\frac{T_C}{\tau}}
    \label{eq:Short-term stab}
\end{equation}

\noindent where $C$ is the fringe contrast, $\Delta f$ is the signal linewidth ($\approx1/2T_R$), $f_0$ is the central frequency (6.8346...~GHz), $T_C$ is the full experimental cycle time ($T_C=140$ms) and $\tau$ is the averaging time. We use this relationship as a basis to optimise our experimental cycle to maximise the potential stability of our clock. A plot of the predicted stability, using the measured SNR, as a function of the Ramsey time is shown in Fig.\ref{fig:SNR_optimisation}. From this we find that as the Ramsey time is increased the predicted stability also improves up to the level of $8.95\times10^{-12}$ for a 10~ms Ramsey time. After this time however the stability begins to decrease as the SNR is degraded. This degradation in SNR is attributed to atomic losses due to both the thermal expansion of the cloud and the atoms falling out of the probe region under gravity. The extension of this Ramsey time should be possible by moving the holes in the cavity body lower down, introducing elliptical holes to maintain good probe-atom overlap along the path of gravity or by implementing a grating-chip atomic fountain\cite{Lewis2022_grating_fountain}. This last option is particularly attractive because as with traditional atomic fountains it would be possible apply both $\pi/2$ pulses when the atoms are at the same vertical point of the cavity. This will allow the phase difference observed by the atoms between the two microwave pulses to be minimised, essential for high contrast Ramsey fringes in a relatively low-Q cavity such as used here. A grating-chip atomic fountain (using CPT interrogation) such as this has already demonstrated Ramsey times out to 100~ms, with a corresponding fringe linewidth of 5~Hz\cite{Lewis2022_grating_fountain}.

%%%% Don't have 1/C factor in this stability estimate yet...

 \begin{figure}[b]
 \includegraphics[width=0.49 \textwidth]{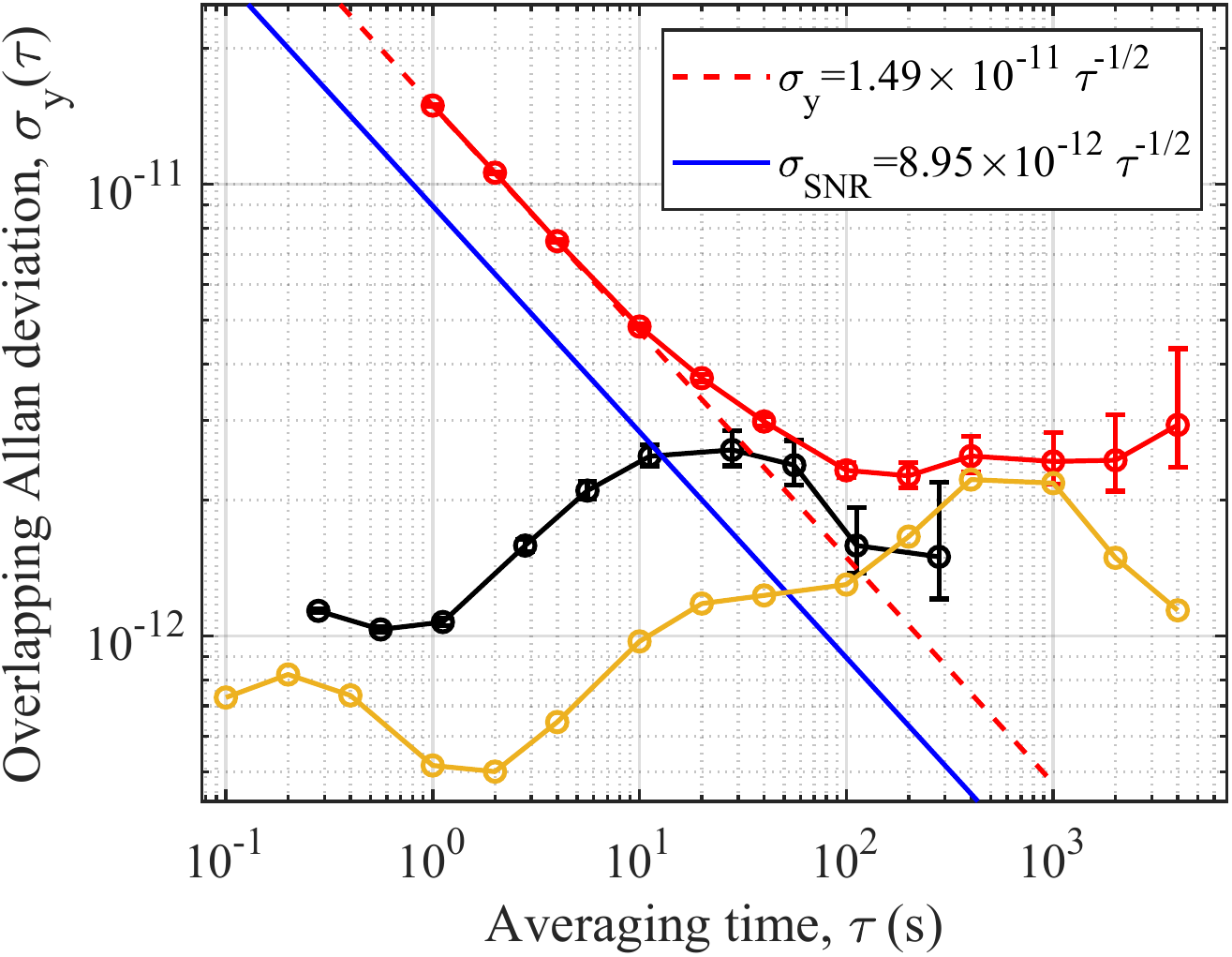}%
 \caption{Red points: Overlapping Allan deviation of local oscillator's stability when locked to atomic signal. Blue solid line: stability predicted by (\ref{eq:Short-term stab}) and measured fringe SNR. Red dashed line: $\tau^{-1/2}$ dependence of the measured 1~s stability. Black points: stability limit due to the second-order Zeeman shift. Yellow points: measured GPSDO reference oscillator stability. Error bars represent 1$\sigma$ confidence bound.}
 \label{fig:ADEV}
 \end{figure}

To assess the stability of our system we stabilise the local oscillator to the atomic signal. The signal is modulated and demodulated at the fringe's half maximum to construct the error signal which is used to feedback onto the local oscillator by voltage tuning its 10~MHz internal reference. Fig.\ref{fig:ADEV} shows an overlapping Allan deviation of the resulting frequency stability when compared to an oven controlled quartz crystal oscillator (Wenzel Associates 501-29647) disciplined to GPS (GPSDO). From Fig.\ref{fig:ADEV} we see that the clock stability averages down with a $1.5\times10^{-11}~\tau^{-1/2}$ dependence out to 10~s, at which point it deviates slightly from the ideal $\tau^{-1/2}$ dependence. This is in reasonable agreement with the theoretical stability obtained from Fig.\ref{fig:SNR_optimisation} at a 10~ms Ramsey time and is well above the ultimate stability limit set by the quantum projection noise (QPN) of $4.9\times 10^{-13}~\tau^{-1/2}$, calculated by replacing the SNR term in (\ref{eq:Short-term stab}) by $\sqrt{N}$, where $N$ is the atom number.

As the experiment is currently operated in a magnetically unshielded environment the instability contribution due to the second-order Zeeman shift was expected to limit the clock stability in the medium-term. This was confirmed by measuring the magnetic field stability via the $\ket{F=1, m_F=1} \rightarrow \ket{F'=2, m_F'=1}$ microwave transition, exhibiting a magnetic field sensitivity of $\beta_1$=1.4~MHz/G \cite{Steck}. The expected stability of the $m_F=0$ clock transition ($\beta_0$=575~Hz/G\textsuperscript{2} ~\cite{Steck}) could then be calculated, shown as black points in Fig.\ref{fig:ADEV}. This plot shows that the second-order Zeeman shift is indeed an important limitation to the clock stability at $\tau>$10~s, due to a pronounced hump in the magnetic field stability around this time, after which the stability averages down slightly to the level of $<2\times10^{-12}$. For future iterations of this set-up it will be imperative to introduce magnetic shielding for long-term stability performance. %\textbf{maybe mention triggering off mains helps but is not solution}. 

Another limiting factor to the clock stability in the medium term is the GPSDO reference. Three-corner-hat stability measurements of this reference against two commercial Cs beam clocks (OSA 3235B and Microchip 5071A) indicates excellent short-term performance of the GPSDO $<1\times 10^{-12}$. At 700~s however, a peak in the instability is observed at $\approx2\times10^{-12}$. This peak is associated with the time constant of the tuning loop used to reference the oscillator to the GPS signal. When taken in conjunction with the stability limit due to the Zeeman effect, these two effects help explain the flattening of the clock stability for averaging times >100~s.

To more fully characterise the short-term stability ($\tau$=1~s) of the system a stability budget of the main instability contributions to the clock was constructed, shown in Table \ref{tab:noise}. The short-term clock performance is found to be primarily limited by the Ramsey fringe SNR, discussed above. In future we expect to be able to improve the SNR contribution to short-term stability towards the projection noise limit. This will be achieved by increasing the Ramsey time, optimisation of the optical detection process and suppression of intensity and frequency fluctuations in the probe beam.

The next largest contributor to the 1~s clock stability after the fringe SNR is electronic noise on the voltage line used to tune the 10~MHz local oscillator. While this is not a limiting factor to the clock stability at present, careful minimisation of this noise source will be required when moving towards a magnetically shielded experiment where the expected clock stability should improve below the $1\times10^{-12}$ level.
The total predicted short term stability from Table \ref{tab:noise} is found to be ~1.6 times better than the measured stability. Efforts are ongoing to identify the remaining instability contributions in order to minimise them in future iterations of the set-up.

%\begin{equation}
%\begin{split}
   % \sigma_{RIN}(\tau) &= \frac{2(1-C/2)}{\pi CQ}  \\ 
   % & \Bigg(\sum_{\substack{k>0 \\ k~odd}}^ \infty sinc^2 \bigg(\frac{k\pi t_d}%{2}T_C\bigg)S_{RIN}\bigg(\frac{k}{2T_C}\bigg)   \Bigg)
%    \end{split}
%\end{equation}

%\begin{equation}
%    \sigma_{RIN}(\tau)= \frac{2(1-C/2)}{\pi CQ} \Bigg(\sum_{\substack{k>0 \\ k~odd}}^ \infty sinc^2 \bigg(\frac{k\pi t_d}{2}T_C\bigg)S_{RIN}\bigg(\frac{k}{2T_C}\bigg)   \Bigg)
%\end{equation}

\begin{table}
\begin{tabular}{llc}
\hline
Noise source                                 &  & \multicolumn{1}{l}{$\sigma$ contribution $(\tau=1s)$} \\ \hline

SNR                                          &  & $8.95 \times10^{-12}$                        \\

Electronic noise                                  &  & $1.14 \times10^{-12}$  
          \\

Dick                                         &  & $6.12 \times10^{-13}$                        \\

Zeeman                                       &  & $1.07 \times10^{-12}$                        \\

%Laser RIN                                    &  & $1.08 \times10^{-13}$                    \\

QPN                                          &  & $4.90 \times10^{-13}$

%    \\
%RIN                                          &  & $1.90 \times10^{-16}$                        
                      \\ \hline
                     
Total \bigg($\sqrt{\sigma_{SNR}^2+\sigma_{LO-tune}^2+\sigma_{Dick}^2}$\bigg) &  & $9.04\times10^{-12}$                         \\ \hline
%Total $\sqrt{\Sigma_i \sigma_{yi}^2 (\tau)}$ &  & $9.12\times10^{-13}$          %               \\ \hline

Measured                                     &  & $1.49\times10^{-11}$                         \\ \hline
\end{tabular}
\caption{Table of noise sources and their contributions to the 1~s Allan deviations stability.}
\label{tab:noise}
\end{table}

%\section{Conclusion}

In conclusion, we have demonstrated a compact cold-atom microwave clock based on an additively manufactured loop-gap-resonator cavity interacting with laser cooled atoms loaded from a GMOT chip. In the present system the experimentally optimised short-term clock stability is measured as $\sigma_y(\tau)=1.5\times10^{-11}~\tau^{-1/2}$, primarily limited by the Ramsey fringe SNR. In the medium-term the clock stability is primarily limited by the second-order Zeeman shift due to operating in an unshilded environment. A secondary limit is placed on the medium-term stability performance by the reference oscillator itself exhibiting a stability bump at around 700~s. By addressing these issues, we expect to improve the short-term stability to be more in-line with other cold-atom cavity clock demonstrations \cite{HORACE_Esnault2010,MuClock,Spectra_dynamics,cold_Atom_cavity_Lee2021}. 

 The current physics package is inherently compact and remains highly amenable to further miniaturisation. The use of additive manufacturing also maintains a highly scalable manufacturing process. Additively manufactured vacuum chambers have also recently demonstrated compatibility with UHV \cite{3d_vacuum_COOPER2021} pressure levels. This raises the enticing possibility of the cavity body itself simultaneously acting as a UHV chamber, allowing the entire physics package to be manufactured as a single bulk component, drastically reducing the size of the system. In addition to this, passively pumped vacuum chambers have now been shown to maintain the UHV levels required for atom trapping for extended periods \cite{passive_vacuum_Little2021,passive_vacuum_Burrow2021}, allowing for a reduction in power consumption by negating the continuous use of an ion pump.
While the stability of our clock is currently several orders of magnitude below the state-of-the-art, the scope for improved performance and further miniaturisation is very good. We therefore believe this work represents a step towards highly compact and portable cold-atom frequency standards.

% If in two-column mode, this environment will change to single-column format so that long equations can be displayed. 
%\begin{widetext}
%$$\mbox{put long equation here}$$
%\end{widetext}

% Figures should be put into the text as floats. 
% Use the graphics or graphicx packages (distributed with LaTeX2e).
% Put the label that you will use with \ref{} command in the braces of the \label{} command.
%
% \begin{figure}
% \end{figure}

% \begin{table}
% \end{tabular}
% \end{table}

\begin{acknowledgments}
The authors would like to thank R. Elvin and J. P. McGilligan  for useful conversations and J. P. McGilligan for careful reading of the manuscript. A.B. was supported by a Ph.D. studentship from the Defence Science and Technology Laboratory (Dstl). E. B., C.A. and  G. M. acknowledge funding from the European space Agency (ESA) and the Swiss Space Office (Swiss Confederation). We gratefully acknowledge funding from the International Network for Microfabrication of Atomic Quantum Sensors (EPSRC EP/W026929/1).

%acknowledgments}

\end{acknowledgments}
% Create the reference section using BibTeX:#
%\newpage

\section*{Data Availability Statement}
The data that support the findings of this study are available from {https://doi.org/10.15129/bb0d3614-6394-4d0e-8649-b1bf66c71722}

\bibliography{ref.bib}

\end{document}